\documentclass[12pt]{iopart}
\usepackage{bm}
\usepackage{graphicx}
\usepackage{cite}

\begin{document}

\title{Topological lattice using multi-frequency radiation}

\author{Tomas Andrijauskas$^1$, I. B. Spielman$^{2,3}$, and Gediminas Juzeli\={u}nas$^3$}

\address{$^1$ Institute of Theoretical Physics and Astronomy, Vilnius University,
Saul\.{e}tekio 3, LT-10257 Vilnius, Lithuania}

\address{$^2$ Joint Quantum Institute, University of Maryland, College Park, Maryland
20742-4111, USA}

\address{$^3$ National Institute of Standards and Technology, Gaithersburg, Maryland
20899, USA}

\ead{gediminas.juzeliunas@tfai.vu.lt}

\begin{abstract}
We describe a novel technique for creating an artificial magnetic
field for ultra-cold atoms using a periodically pulsed pair of counter
propagating Raman lasers that drive transitions between a pair of
internal atomic spin states: a multi-frequency coupling term. In conjunction
with a magnetic field gradient, this dynamically generates a rectangular
lattice with a non-staggered magnetic flux. For a wide range of parameters,
the resulting Bloch bands have non-trivial topology, reminiscent of
Landau levels, as quantified by their Chern numbers. 
\end{abstract}
\maketitle

\section{Introduction}

Ultracold atoms find wide applications in realising condensed matter
phenomena \cite{Greiner2002,Lewenstein2007,Bloch2008a,Lewenstein2012}.
Since ultracold atom systems are ensembles of electrically neutral
atoms, various methods have been used to simulate Lorentz-type forces,
with an eye for realizing physics such as the quantum Hall effect
(QHE). Lorentz forces are present in spatially rotating systems \cite{Matthews1999a,Madison2001,Abo-Shaeer2001,Cooper2008,Fetter2009,Gemelke2010,Wright13PRL}
and appear in light-induced geometric potentials \cite{Dalibard2011,Goldman2014}.
The magnetic fluxes achieved with these methods are not sufficiently
large for realizing the integer or fractional QHE. In optical lattices,
larger magnetic fluxes can be created by shaking the lattice potential
\cite{struck12,Windpassinger2013RPP,Jotzu2014,Eckardt16review}, combining
static optical lattices along with laser-assisted spin or pseudo spin
coupling \cite{Javanainen2003,Jaksch2003,Osterloh2005,Dalibard2011,Cooper2011PRL,Aidelsburger:2013,Goldman2014,goldman16review,miyake13harper};
current realizations of these techniques are beset with micro motion
and interaction induced heating effects. Here we propose a new method
that simultaneously creates large artificial magnetic fields and a
lattice that may overcome these limitations.

Our technique relies on a pulsed atom-light coupling between internal
atomic states along with a state-dependent gradient potential that
together create a two-dimensional (2D) periodic potential with an
intrinsic artificial magnetic field. With no pre-existing lattice
potential, there are no a priori resonant conditions that would otherwise
constrain the modulation frequency to avoid transitions between original
Bloch bands \cite{Weinberg15PRA}. For a wide range of parameters,
the ground and excited bands of our lattice are topological, with
nonzero Chern number. Moreover, like Landau levels the lowest several
bands can all have unit Chern number.

The manuscript is organized as follows. Firstly, we describe a representative
experimental implementation of our technique directly suitable for
alkali atoms. Secondly, because the pulsed atom-light coupling is
time-periodic, we use Floquet methods to solve this problem. Specifically,
we employ a stroboscopic technique to obtain an effective Hamiltonian.
Thirdly, using the resulting band structure we obtain a phase diagram
which includes a region of Landau level-like bands each with unit
Chern number.

\section{Pulsed lattice}

\begin{figure}
\begin{center}
\includegraphics{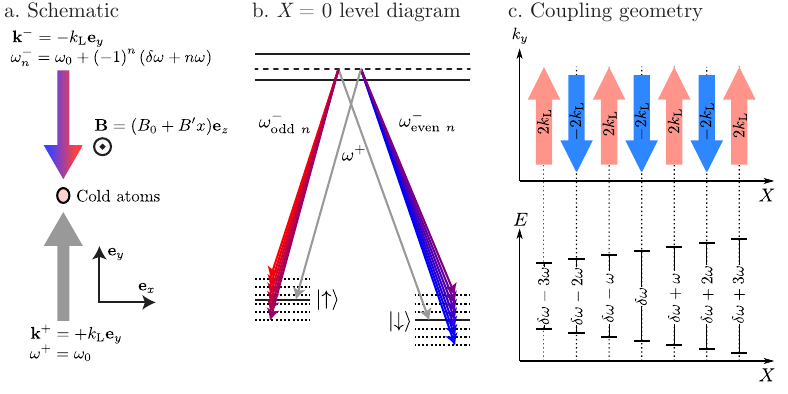}
\end{center}
\caption{Floquet flux lattice. a. Experimental schematic depicting a cold cloud
of atoms in a gradient magnetic field, illuminated by a pair of counter-propagating
laser beams tuned near two-photon Raman resonance. The down-going
beam includes sidebands both to the red and blue of the carrier ($\omega_{0}$)
in resonance at different spatial positions along ${\bf e}_{x}$.
b. Level diagram showing even and odd side-bands linking the $\left|\uparrow\right\rangle $
and $\left|\downarrow\right\rangle $ states with differing detuning
from resonance at $X=0$. c. Spatially dependent coupling. Bottom:
different frequency components are in two-photon resonance in different
$X$ positions. Top: the recoil kick associated with the Raman transition
is along $\pm\mathbf{e}_{y}$ and thus alternates spatially depending
on whether the Raman transition is driven from the red or blue sideband
of the down-going laser beam.}
\label{fig:schematic} 
\end{figure}

Figure 1 depicts a representative experimental realization of the
proposed method. A system of ultracold atoms is subjected to a magnetic
field with a strength $B(X)=B_{0}+B^{\prime}X$. This induces a position-dependent
splitting $g_{F}\mu_{{\rm B}}B$ between the spin up and down states;
$g_{F}$ is the Landé $g$-factor and $\mu_{{\rm B}}$ is the Bohr
magneton. Additionally, the atoms are illuminated by a pair of Raman
lasers counter propagating along ${\bf e}_{y}$, i.e. perpendicular
to the detuning gradient. The first beam (up-going in Fig.~\ref{fig:schematic}(a))
is at frequency $\omega^{+}=\omega_{0}$, while the second (down-going
in Fig.~\ref{fig:schematic}(a)) contains frequency components $\omega_{n}^{-}=\omega_{0}+(-1)^{n}(\delta\omega+n\omega)$;
the difference frequency between these beams contains frequency combs
centered at $\pm\delta\omega$ with comb teeth spaced by $2\omega$,
as shown in Fig.~\ref{fig:schematic}(b). In our proposal, the Raman
lasers are tuned to be in nominal two-photon resonance with the Zeeman
splitting from the large offset field $B_{0}$ such that $g_{F}\mu_{{\rm B}}B_{0}=\hbar\delta\omega_{0}$,
making the frequency difference $\omega_{n=0}^{-}-\omega^{+}$ resonant
at $X=0$, where $B=B_{0}$. Intuitively, each additional frequency
component $\omega_{n}^{-}$ adds a resonance condition at the regularly
spaced points $X_{n}=n\hbar\omega/g_{F}\mu_{{\rm B}}B^{\prime}$,
however, transitions using even-$n$ side bands give a recoil kick
opposite from those using odd-$n$ side bands (see Fig.~\ref{fig:schematic}(c)).
Each of these coupling-locations locally realizes synthetic magnetic
field experiment performed at the  National Institute
of Standards and Technology (NIST)~\cite{Lin2009b}, arrayed in a
manner to give a rectified artificial magnetic field with a non-zero
average that we will show is a novel flux lattice.

In practice only a finite number of lattice teeth are
needed, owing to the finite spatial extent of a trapped atomic gas.
In rough numbers the spatial extent of a quantum degenerate gas is
about $20\mu m$, and if we select a very large gradient corresponding
to a lattice spacing of $0.5\mu m$, this gives just 40 comb teeth.
Note also that generating the frequency comb is a very straightforward
procedure. In the laboratory one uses acoustic-optical modulators
(AOMs) to frequency shift laser beams by a frequency defined by a
laboratory radiofrequency (rf) source. Therefore creating a
comb is a simple matter of first creating a frequency comb -- 
simple with rf -- and then feeding that signal into the
AOM. This sort of frequency synthesis is being carried in a routine
manner in the ultracold atom labs.

We formally describe our system by first making the rotating wave
approximation (RWA) with respect to the large offset frequency $\omega_{0}$.
This situation is modeled in terms of a spin-1/2 atom of mass $M$
and wave-vector  $\bm{K}=-i\boldsymbol{\nabla}$ with
a Hamiltonian 
\begin{equation}
H(t)=H_{0}+V(t).\label{eq:full-Hamiltonian}
\end{equation}
The first term is 
\begin{equation}
H_{0}=\frac{\hbar^{2}\bm{K}^{2}}{2M}+\frac{\Delta(X)}{2}\sigma_{3},\label{eq:Hamiltonian0}
\end{equation}
where $\Delta(X)=\Delta^{\prime}X$ describes the detuning gradient
along $\mathbf{e}_{x}$ axis, and $\sigma_{3}=|\!\uparrow\rangle\langle\uparrow\!|-|\!\downarrow\rangle\langle\downarrow\!|$
is a Pauli spin operator. In the RWA only near-resonant terms are
retained, giving the Raman coupling described by 
\begin{equation}
V(t)=V_{0}\sum_{n}\left[{\rm e}^{{\rm i}(K_{0}Y-2n\omega t)}+{\rm e}^{{\rm i}(-K_{0}Y-(2n+1)\omega t)}\right]|\!\downarrow\rangle\langle\uparrow\!|+{\rm H.\,c.}\,.\label{eq:Raman-coupling}
\end{equation}
The first term describes coupling from the sidebands with even frequencies
$2n\omega$, whereas the second term describes coupling from the sidebands
with odd frequencies $\left(2n+1\right)\omega$. The recoil kick is
aligned along $\pm\mathbf{e}_{y}$ with opposite sign for the even
and odd frequency components. In writing Eq.(\ref{eq:Raman-coupling})
we assumed that the coupling amplitude $V_{0}$ and the associated
recoil wave number $K_{0}$ are the same for all frequency components.
The coupling Hamiltonian $V(t)$ and therefore the full Hamiltonian
$H(t)$ are time-periodic with period $2\pi/\omega$, and we accordingly
apply Floquet techniques.

\section{Theoretical analysis}

The outline of this Section is as follows. (1) We begin the analysis
of the Hamiltonian given by Eq.~(\ref{eq:full-Hamiltonian}) by moving
to dimensionless units; (2) subsequently derive an approximate effective
Hamiltonian from the single-period time evolution operator; (3) provide
an intuitive description in terms of adiabatic potentials; and (4)
finally solve the band structure, evaluate its topology and discuss
possibilities of the experimental implementation.

\subsection{Dimensionless units}

For the remainder of the manuscript we will use dimensionless units.
All energies will be expressed in units of $\hbar\omega$, derived
from the Floquet frequency $\omega$; time will be expressed in units
of inverse driving frequency $\omega^{-1}$, denoted by $\tau=\omega t$;
spatial coordinates will be expressed in units of inverse recoil momentum
$K_{0}^{-1}$, denoted by lowercase letters $(x,y)=K_{0}(X,Y)$. In
these units, the Hamiltonian (\ref{eq:full-Hamiltonian}) takes the
form 
\begin{equation}
h(\tau)=\frac{H(\tau/\omega)}{\hbar\omega}=E_{\rm{r}}\bm{k}^{2}+\frac{1}{2}\boldsymbol{\Omega}(\tau)\cdot\boldsymbol{\sigma}\,,\label{eq:dimless-Hamiltonian}
\end{equation}
where $E_{\rm{r}}=\hbar^{2}K_{0}^{2}/(2M\hbar\omega)$ is the dimensionless
recoil energy associated with the recoil wavenumber $K_{0}$; $\bm{k}=\bm{K}/K_{0}$
is the dimensionless wavenumber. The dimensionless coupling 
\begin{equation}
\boldsymbol{\Omega}(x,y,\tau)=\left(2{\rm Re}\,u(y,\tau),\,2{\rm Im}\,u(y,\tau),\,\beta x\right)\label{eq:full-coupling}
\end{equation}
includes a combination of position-dependent detuning and Raman coupling.
Here $\beta=\Delta^{\prime}/(\hbar\omega k_{0})$ describes the linearly
varying detuning in dimensionless units; the function $u(y,\tau)=v_{0}\sum_{n}\left\{ \exp[{\rm i}(y-2n\tau)]+\exp[{\rm i}(-y-(2n+1)\tau)]\right\} $
is a dimensionless version of the sum in Eq.~(\ref{eq:Raman-coupling})
with $v_{0}=V_{0}/(\hbar\omega)$.

In the time domain the coupling given by Eq. (\ref{eq:full-coupling})
is 
\begin{equation}
\frac{1}{2}\boldsymbol{\Omega}(\tau)\cdot\boldsymbol{\sigma}=\frac{1}{2}\beta x\sigma_{3}+\sum_{l}v_{l}(y)\delta(\tau-\pi l),\label{eq:coupling-even-odd}
\end{equation}
with 
\begin{equation}
v_{l}(y)=\pi v_{0}\left[{\rm e}^{{\rm i}y}+(-1)^{l}{\rm e}^{-{\rm i}y}\right]|\!\downarrow\rangle\langle\uparrow\!|+{\rm H.\,c.}\,.\label{eq:v_l}
\end{equation}
In this way we have separated the spatial and temporal dependencies
in the coupling (\ref{eq:coupling-even-odd}).

\subsection{Effective Hamiltonian}

We continue our analysis by deriving an approximate Hamiltonian that
describes the complete time evolution over a single period from $\tau=0-\epsilon$
to $\tau=2\pi-\epsilon$ with $\epsilon\to0$. This evolution includes
a kick $v_{0}$ at the beginning of the period $\tau_{+}=0$ and a
second kick $v_{1}$ in the middle of the period $\tau_{-}=\pi$;
between the kicks the evolution includes the kinetic and gradient
energies. In the full time period, the complete evolution operator
is a product of four terms: 
\begin{equation}
U(2\pi,0)\equiv\lim_{\epsilon\to0}U(2\pi-\epsilon,0-\epsilon)=U_{0}U_{\rm{kick}}^{(1)}U_{0}U_{\rm{kick}}^{(0)}.\label{eq:time-evolution}
\end{equation}
Here 
\begin{equation}
U_{0}=\exp\left\{ -{\rm i}\pi\left[E_{\rm{r}}\bm{k}^{2}+\frac{1}{2}\sigma_{3}\beta x\right]\right\} \label{eq:time-evolution-0}
\end{equation}
is the evolution operator over a half period, generated by kinetic
energy and gradient. The operator 
\begin{equation}
U_{\rm{kick}}^{(l)}=\exp\left[-{\rm i}v_{l}(y)\right].\label{eq:kicks}
\end{equation}
describes a kick at $\tau=l\pi$.

We obtain an effective Hamiltonian by assuming that the Floquet frequency
$\omega$ greatly exceeds the recoil frequency, $1\gg E_{\rm{r}}$,
allowing us to ignore the commutators between the kinetic energy and
functions of coordinates in eq.(\ref{eq:time-evolution}). We then
rearrange terms in the full time evolution operator (\ref{eq:time-evolution})
and obtain (see Appendix A) 
\begin{equation}
U_{\rm{eff}}=\exp\left\{ -{\rm i}2\pi\left[E_{\rm{r}}\bm{k}^{2}+v_{\rm{eff}}\right]\right\} ,\label{eq:effective-time-evolution-op}
\end{equation}
where $v_{\rm{eff}}$ is an effective coupling defined by 
\begin{equation}
\exp\left(-{\rm i}2\pi v_{\rm{eff}}\right)={\rm e}^{-{\rm i}\pi\sigma_{3}\beta x/2}U_{\rm{kick}}^{(1)}{\rm e}^{-{\rm i}\pi\sigma_{3}\beta x/2}U_{\rm{kick}}^{(0)}.\label{eq:v-eff-exponent}
\end{equation}

The function $v_{l}(y)$ entering the kick operators $U_{\rm{kick}}^{(l)}$
is spatially periodic along the $y$ direction with a period $2\pi$.
This period can be halved to $\pi$ by virtue of a gauge transformation
$U=\exp(-{\rm i}y\sigma_{3}/2)$. Subsequently, when exploring energy
bands and their topological properties, this prevents problems arising
from using a twice larger elementary cell. Following this transformation
the evolution operator becomes 
\begin{equation}
U_{\rm{eff}}=\exp\left\{ -{\rm i}2\pi\left[E_{\rm{r}}\left(\bm{k}+\sigma_{3}\bm{e}_{y}/2\right)^{2}+v_{\rm{eff}}\right]\right\} ,\label{eq:effective-time-evolution-op-1}
\end{equation}
where $v_{l}(y)$ featured in the kick operators $U_{\rm{kick}}^{(l)}$
has now the spatial periodicity $\pi$ along the $y$ direction, i.e.
it should be replace to 
\begin{equation}
v_{l}(y)=\pi v_{0}\left[{\rm e}^{{\rm i}2y}+(-1)^{l}\right]|\!\downarrow\rangle\langle\uparrow\!|+{\rm H.\,c.}\,.\label{eq:v_l-1}
\end{equation}

The algebra of Pauli matrices allows us to write the effective coupling
 $v_{\rm{eff}}(\bm{r})$ featured in the evolution equations
(\ref{eq:v-eff-exponent})-(\ref{eq:effective-time-evolution-op-1})
as: 
\begin{equation}
v_{\rm{eff}}(\bm{r})=\frac{1}{2}\bm{\Omega}_{\rm{eff}}(\bm{r})\cdot\bm{\sigma},\label{eq:effective-coupling}
\end{equation}
where $\bm{\Omega}_{\rm{eff}}=\left(\Omega_{\rm{eff},1},\Omega_{\rm{eff},2},\Omega_{\rm{eff},3}\right)$
is a position-dependent effective Zeeman field which takes the analytic
form 
\begin{equation}
\exp\left(-{\rm i}2\pi v_{\rm{eff}}\right)=q_{0}-{\rm i}q_{1}\sigma_{1}-{\rm i}q_{2}\sigma_{2}-{\rm i}q_{3}\sigma_{3}.\label{eq:V-eff-exponent-in-q}
\end{equation}
Here $q_{0}$, $q_{1}$, $q_{2}$ and $q_{3}$ are real functions
of the coordinates $(x,y)$, allowing to express the effective Zeeman
field as 
\begin{equation}
\bm{\Omega}_{\rm{eff}}=\pi^{-1}\frac{\bm{q}}{||\bm{q}||}\arccos q_{0},\label{eq:effective-magnetic-field}
\end{equation}
where $\bm{q}$ is a shorthand of a three dimensional vector $(q_{1},q_{2,}q_{3})$.
In general the equation (\ref{eq:V-eff-exponent-in-q}) gives multiple
solutions that correspond for different Floquet bands. Our choice
(\ref{eq:effective-magnetic-field}) picks only to the two bands that
lie in the energy window from $-1/2$ to $1/2$ covering a single
Floquet period.

Comparing (\ref{eq:v-eff-exponent}) and (\ref{eq:V-eff-exponent-in-q})
and multiplying four matrix exponents give explicit expressions 
\begin{eqnarray}
q_{0} & =\cos f_{1}\cos f_{2}\cos(\pi\beta x),\label{eq:function-q0}\\
q_{1} & =\sin f_{1}\cos f_{2}\cos(y+\pi\beta x)-\cos f_{1}\sin f_{2}\sin(y),\label{eq:function-q1}\\
q_{2} & =\sin f_{1}\cos f_{2}\sin(y+\pi\beta x)+\cos f_{1}\sin f_{2}\cos(y),\label{eq:function-q2}\\
q_{3} & =\cos f_{1}\cos f_{2}\sin(\pi\beta x)-\sin f_{1}\sin f_{2}\label{eq:function-q3}
\end{eqnarray}
with 
\begin{eqnarray}
f_{1}(y) & =2\pi v_{0}\cos(y),\label{eq:function-f1}\\
f_{2}(y) & =2\pi v_{0}\sin(y).\label{eq:function-f2}
\end{eqnarray}

\begin{figure}
\begin{center}
\includegraphics[width=1.75in]{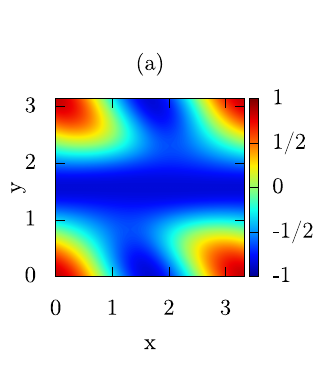}
\includegraphics[width=1.75in]{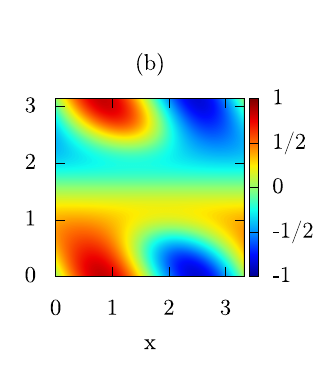}
\includegraphics[width=1.75in]{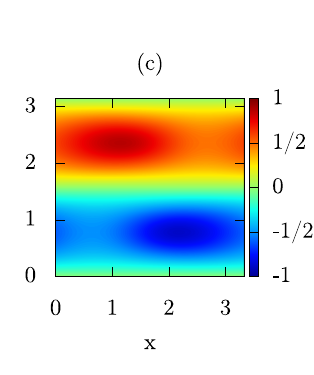}
\end{center}
\caption{Coupling components (a) $\Omega_{\rm{eff},1}(\bm{r})$, (b) $\Omega_{\rm{eff},2}(\bm{r})$
and (c) $\Omega_{\rm{eff},3}(\bm{r})$ for $v_{0}=0.25$ and $\beta=0.6$
 calculated using Eqs.~(\ref{eq:effective-magnetic-field})-(\ref{eq:function-f2}).
The corresponding eigenvalues of the coupling $v_{\pm}(\bm{r})=\pm\Omega_{\rm{eff}}/2$
are presented by the thick red solid lines in the fig.~\ref{fig:floquet-spectrum}.}
\label{fig:coupling} 
\end{figure}

These explicit expressions show that the resulting effective Zeeman
field (\ref{eq:effective-magnetic-field}) and the associated effective
coupling (\ref{eq:effective-coupling}) are periodic along both $\bm{e}_{x}$
and $\bm{e}_{y}$, with spatial periods $a_{x}=2/\beta$ and $a_{y}=\pi$
respectively. Therefore, although the original Hamiltonian containing
the spin-dependent potential slope $\propto x\sigma_{3}$ is not periodic
along the $x$ direction, the effective Floquet Hamiltonian is. The
spatial dependence of the Zeeman field components $\Omega_{\rm{eff},1}$,
$\Omega_{\rm{eff},2}$ and $\Omega_{\rm{eff},3}$ is presented
in the fig.~\ref{fig:coupling} for $\beta=0.6$ giving an approximately
square unit cell. In fig.~\ref{fig:coupling} we select $v_{0}=0.25$
where the absolute value of the Zeeman field $\Omega_{\rm{eff}}$
is almost uniform, as is apparent from the nearly flat adiabatic bands
shown in fig.~\ref{fig:floquet-spectrum} below.

\subsection{Adiabatic evolution and magnetic flux\label{subsec:Adiabatic-evolution-and}}

\begin{figure}
\begin{center}
\includegraphics[width=4in]{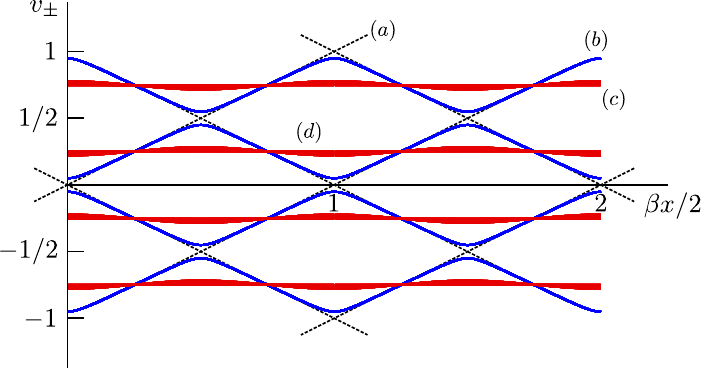}
\end{center}

\caption{Adiabatic Floquet potentials  given by Eq.~(\ref{eq:eigenvalues-of-V-eff})
for $\beta=0.6$ us. (a) Thin black dotted lines denote the spin-dependent
gradient slopes without including the Raman coupling ($v_{0}=0$);
(b) thin blue solid lines denote effective adiabatic potentials for
weak Raman coupling ($v_{0}=0.05$) (c) red solid lines denote nearly
flat adiabatic potentials that are achieved for stronger Raman coupling
($v_{0}=0.25$). All the curves are projected into $x$ plane for
various $y$ values. A weak $y$ dependence of the adiabatic potentials
is seen to appear in the strong coupling case (c) making the superimposed
red lines thicker.}
\label{fig:floquet-spectrum} 
\end{figure}

Before moving further to an explicit numerical analysis of the band
structure, we develop an intuitive understanding by performing an
adiabatic analysis of motion governed by effective Hamiltonian 
\begin{equation}
h_{\rm{eff}}(\bm{r})=E_{\rm{r}}\left(\bm{k}+\sigma_{3}\bm{e}_{y}/2\right)^{2}+\frac{1}{2}\bm{\Omega}_{\rm{eff}}\cdot\bm{\sigma}\,\label{eq:h_eff}
\end{equation}
featured in the evolution operator $U_{\rm{eff}}$, Eq.~(\ref{eq:effective-time-evolution-op}).
The coupling field $\bm{\Omega}_{\rm{eff}}(\bm{r})$ is parametrized
by the spherical angles $\theta(\bm{r})$ and $\phi(\bm{r})$ defined
by 
\begin{eqnarray}
\cos\theta & =\frac{\Omega_{\rm{eff},3}}{\Omega_{\rm{eff}}},\label{eq:spherical-cos-theta}\\
\tan\phi & =\frac{\Omega_{\rm{eff},2}}{\Omega_{\rm{eff},1}}.\label{eq:spherical-tan-phi}
\end{eqnarray}
This gives the effective coupling \cite{Dalibard2011} 
\begin{equation}
\frac{1}{2}\bm{\Omega}_{\rm{eff}}\cdot\bm{\sigma}=\frac{1}{2}\Omega_{\rm{eff}}\left[\begin{array}{cc}
\cos\theta & {\rm e}^{-{\rm i}\phi}\sin\theta\\
{\rm e}^{{\rm i}\phi}\sin\theta & -\cos\theta
\end{array}\right]\,,\label{eq:eff-coupling-in-spherical-coords}
\end{equation}
characterized by the position-dependent eigenstates 
\begin{equation}
\left|+\right\rangle =\left(\begin{array}{c}
\cos\left(\theta/2\right)\\
{\rm e}^{{\rm i}\phi}\sin\left(\theta/2\right)
\end{array}\right)\,,\qquad\left|-\right\rangle =\left(\begin{array}{c}
-{\rm e}^{-{\rm i}\phi}\sin\left(\theta/2\right)\\
\cos\left(\theta/2\right)
\end{array}\right)\,.\label{eq:pm-states}
\end{equation}
The corresponding eigenvalues 
\begin{equation}
v_{\pm}(\bm{r})=\pm\frac{1}{2}\Omega_{\rm{eff}},\label{eq:eigenvalues-of-V-eff}
\end{equation}
are shown in Fig.~\ref{fig:floquet-spectrum} for various value of
the Raman coupling $v_{0}$. As one can see in Fig.~\ref{fig:floquet-spectrum},
for $v_{0}=0.25$ the resulting bands $v_{\pm}(\bm{r})$ (adiabatic
potentials) are flat and have a considerable gap $\approx\omega/2$,
a regime suitable for a description in terms of an adiabatic motion
in selected bands \cite{Zoller2008}.

As in Ref.~\cite{Juzeliunas2012}, we consider the adiabatic motion
of the atom in one of these flat adiabatic bands with the projection
Shrödinger equation that includes a geometric vector potential 
\begin{equation}
\bm{A}_{\pm}(\bm{r})=\pm\frac{1}{2}\left(\cos\theta-1\right)\nabla\phi\,.\label{eq:geometric-vector-potential}
\end{equation}
This provides a synthetic magnetic flux density $\bm{B}_{\pm}(\bm{r})=\nabla\times\bm{A}_{\pm}(\bm{r})$.
The geometric vector potential $\bm{A}_{\pm}(\bm{r})$ may contain
Aharonov-Bohm type singularities, that give rise to a synthetic magnetic
flux over an elementary cell 
\begin{equation}
\alpha_{\pm}=-\sum\oint_{{\rm singul}}{\rm d}\bm{r}\cdot\bm{A}_{\pm}(\bm{r}).\label{eq:synthetic-magnetic-flux}
\end{equation}
The singularities appear at points where $\theta=\pi$, where the
angle $\phi$ and its gradient $\nabla\phi$ are undefined and $\cos\theta=-1$.
The term $\cos\theta-1$ in (\ref{eq:geometric-vector-potential})
is non zero and does not remove the undefined phase $\nabla\phi$.
Our unit cell contains two such singularities located at $\bm{r}=(a_{x},3a_{y})/4$
and $\bm{r}=(3a_{x},a_{y})/4$, containing the same flux, so that
they do not compensate each other, giving the synthetic magnetic flux
$\pm2\pi$ in each unit cell. Note that usually the optical lattices
are sufficiently deep, and the $\pm2\pi$ flux per elementary is topologically
trivial. In that case the tight binding model can be applied, with
the tunneling taking place only between the nearest-neighboring sites
of the square plaquette. The $\pm2\pi$ flux over the square plaquette
can then be eliminated by a gauge transformation. Yet if the lattice
is shallow enough, the tight binding model is not applicable and the
above arguments do not work. In the present situation, the most interesting
topological lattice appears for a flat adiabatic trapping potential
shown by a solid red curve in Fig.~\ref{fig:floquet-spectrum}. In
such a situation there are no well defined lattice sites, and the
$\pm2\pi$ flux per elementary cell results in topologically non-trivial
bands explored in the next Subsection. 

\begin{figure}
\begin{center}
\includegraphics[width=3in]{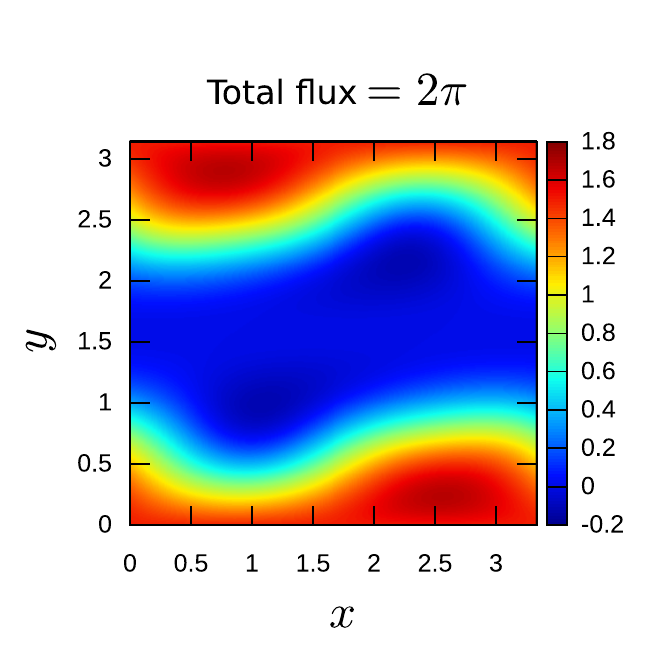}
\end{center}
\caption{Geometric flux density $\bm{B}_{\pm}(\bm{r})=\nabla\times\bm{A}_{\pm}(\bm{r})$
computed for $v_{0}=0.25$ and $\beta=0.6$  using Eq.~(\ref{eq:geometric-vector-potential}).
The overall spatial structure of this flux density does not depend
on the gradient $\beta$; rather it scales with the corresponding
lattice constant $a_{x}=2/\beta$. }
\label{fig:flux-density} 
\end{figure}

For a weak coupling (such as $v=0.05$) the geometric flux density
$\bm{B}(\bm{r})\equiv\bm{B}_{\pm}(\bm{r})$ is concentrated around
the intersection points of the gradient slopes shown in in Fig.~\ref{fig:floquet-spectrum}
and has a very weak $y$ dependence. With increasing the coupling
$v$, the flux extends beyond the intersection areas and acquires
a $y$ dependence. Fig.~\ref{fig:flux-density} shows the geometric
flux density $\bm{B}(\bm{r})\equiv\bm{B}_{+}(\bm{r})$ for the strong
coupling ($v_{0}=0.25$) corresponding to the most flat adiabatic
bands. In this regime the flux develops stripes in the $x$ direction
and has a strong $y$ dependence. For the whole range of coupling
strengths $0\le v_{0}\le1/2$ the total synthetic magnetic flux per
unit cell is $2\pi$ and is independent of the Floquet frequency $\omega$
and the gradient $\beta$.

Now let us discuss the effect of an extra spin-independent
trapping potential. The present scheme requires a large spin-dependent
energy gradient which would have a huge influence on the relative
trapping for the two spin states without the Raman coupling or for
a weak Raman coupling. In that case one would expect that the stable
positions for any trapped sample of the two spin states would live
at entirely distinct locations, possibly with no overlap. Yet we are
interested mostly in a sufficiently strong Raman coupling where the
two spin states get mixed, and the atomic motion takes place in almost
flat adiabatic potentials shown in red in Fig.~\ref{fig:floquet-spectrum}.
Therefore the atoms are no longer affected by the steep spin-dependent
potential slopes, and the spin-independent trapping potential would
not cause separation of different spin states. Instead, the extra
spin-independent parabolic trapping potential would simply make the
flat adiabatic trapping potentials parabolic. Of course, one needs
to be all the time in the regime where the Raman coupling is strong
compared to the characteristic energy of the spin-dependent potential
slope. That is why we propose to introduce the spin-dependent
potential gradient only at the final stage of the adiabatic protocol
discussed in Sect. \ref{subsec:Loading-into-dressed}. 

\subsection{Band structure and Chern numbers}

We analyze the topological properties of this Floquet flux lattice
by explicitly numerically computing the band structure and associated
Chern number using the effective Hamiltonian (\ref{eq:h_eff}) without
making the adiabatic approximation introduced in Sec.~\ref{subsec:Adiabatic-evolution-and}.
Again the gradient of the original magnetic field is such that we
approximately get a square lattice, $\beta=0.6$. Furthermore, we
choose the Floquet frequency to be ten times larger than the recoil
energy, $E_{\rm{r}}=0.1$.  Note that one can alter
the length of the plaquette along the x direction (and thus the flux
density) by changing $\beta$ representing the potential gradient
along the $x$ axis. 

\begin{figure}
\begin{center}
\includegraphics[width=4in]{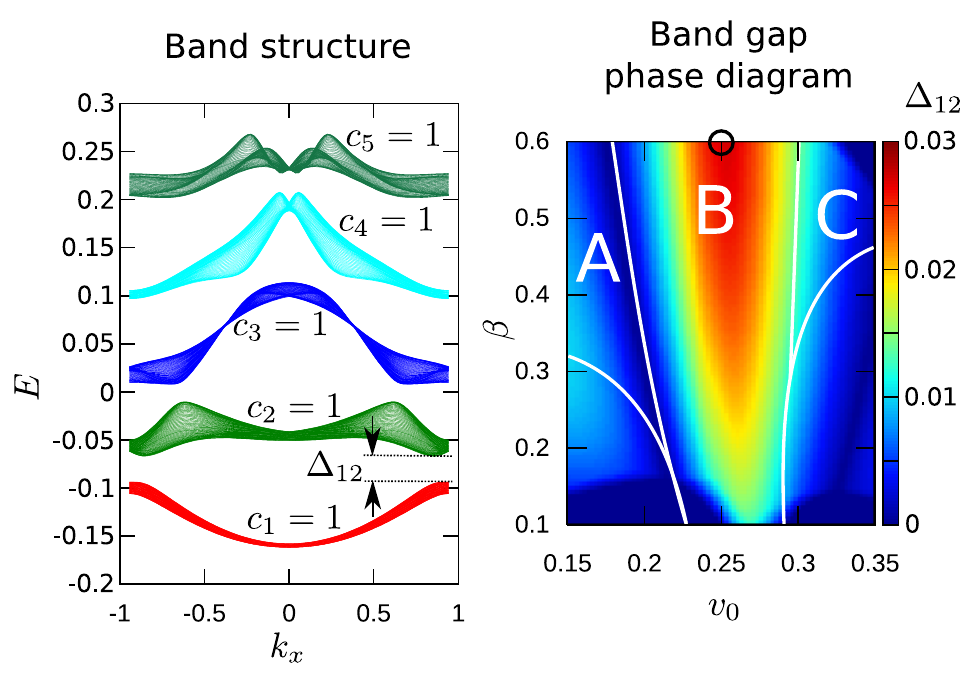}
\end{center}

\caption{Left: band structure  calculated using the effective Hamiltonian
(\ref{eq:h_eff}) for $v_{0}=0.25$, $\beta=0.6$ and $E_{\rm{r}}=0.1$.
Right: The band gap $\Delta_{12}$ between the first and second bands
for $E_{\rm{r}}=0.1$ and various values of $v_{0}$ and $\beta$.}
\label{fig:bands-chern} 
\end{figure}

\begin{figure}
\begin{center}
\includegraphics[width=3in]{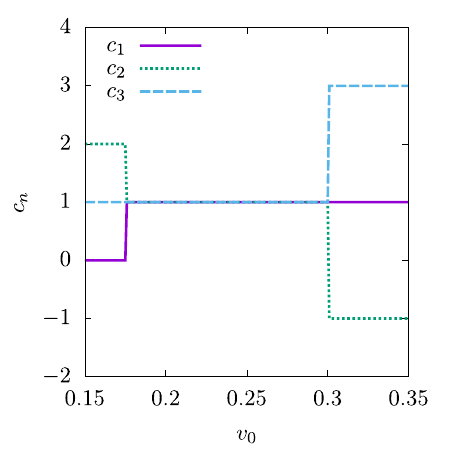}
\end{center}
\caption{  Dependence of Chern number for the bands calculated
using the effective Hamiltonian (\ref{eq:h_eff}) on the coupling
strength $v_{0}$ for $\beta=0.6$ and $E_{\rm{r}}=0.1$. Here we
present the Chern numbers $c_{1}$, $c_{2}$ and $c_{3}$ of the three
lowest bands. }
\label{fig:chern} 
\end{figure}

First, let us consider the case where $v_{0}=0.25$ corresponding
to the most flat adiabatic potential. In this situation the Chern
numbers of the first five bands appear to be equal to the unity, as
one can see in the left part of Fig.~\ref{fig:bands-chern}. Thus
the Hall current should monotonically increase when filling these
bands. This resembles the Quantum Hall effect involving the Landau
levels. Second, we check what happens when we leave the regime $v_{0}=0.25$
where the adiabatic potential is flat, and consider lower and higher
values of the coupling strength $v_{0}$. Near $v_{0}=0.175$ we find
a topological phase transition where the lowest two energy bands touch
and their Chern numbers change to $c_{1}=0$ and $c_{2}=2$, while
the Chern numbers of the higher bands remain unchanged, illustrated
in fig.~\ref{fig:chern}. In a vicinity of $v_{0}=0.3$ there is
another phase transition, where the second and third bands touch,
leading to a new distribution of Chern numbers: $c_{1}=1$, $c_{2}=-1$,
$c_{3}=3$, $c_{4}=1$. Interestingly the Chern numbers of the second
and the third bands jump by two units during such a transition.

Finally, we explore the robustness of the topological bands. The right
part of Fig.~\ref{fig:bands-chern} shows the dependence of the band
gap $\Delta_{12}$ between the first and second bands on the coupling
strength $v_{0}$ and the potential gradient $\beta$. One can see
that the band gap is maximum for $v_{0}=0.25$ when the adiabatic
potential is the most flat. The gap increases by increasing the gradient
$\beta$, simultaneously extending the range of the $v_{0}$ values
where the band gap is nonzero. Therefore to observe the topological
bands, one needs to take a proper value of the Raman coupling $v_{0}\approx0.25$
and a sufficiently large gradient $\beta$, such as $\beta=0.6$.

We now make some numerical estimates to confirm that this scheme is
reasonable. We consider an ensemble of $^{87}{\rm Rb}$ atoms, with
$\left|{\uparrow}\right>=\left|f=2,m_{F}=2\right>$ and $\left|\downarrow\right>=\left|{f=1,m_{F}=1}\right>$;
the relative magnetic moment of these hyperfine states is $\approx2.1\ {\rm MHz}/{\rm G}$,
where $1\ {\rm G}=10^{-4}\ {\rm T}$. For a reasonable magnetic field
gradient of $300\ \mathrm{G}/\mathrm{cm}$, this leads to the $\Delta^{\prime}/\hbar\approx2\pi\times600\ \mathrm{MHz}/\mathrm{cm}=2\pi\times60\ {\rm kHz}/\mu{\rm m}$
detuning gradient. For $^{87}{\rm Rb}$ with $\lambda=790\ {\rm nm}$
laser fields the recoil frequency is $\omega_{r}/2\pi=3.5\ \mathrm{kHz}$.
Along with the driving frequency $\omega=10\omega_{r}$, this provides
the dimensionless energy gradient $\beta=\Delta^{\prime}/(\hbar\omega k_{0})\approx1.3$,
allowing easy access to the topological bands displayed 
in Fig.~\ref{fig:bands-chern}.

\subsection{Loading into dressed states\label{subsec:Loading-into-dressed}}

Adiabatic loading into this lattice can be achieved by extending the
techniques already applied to loading in to Raman dressed states \cite{Lin2009a}.
The loading technique begins with a  Bose-Einstein condensate
(BEC) in the lower energy $\downarrow$ state in a uniform magnetic
field $B_{0}$. Subsequently one slowly ramps on a single off resonance
RF coupling field and the adiabatically ramp the RF field to resonance
(at frequency $\delta\omega$). This RF dressed state can be transformed
into a resonant Raman dressed by ramping on the Raman lasers (with
only the $\omega_{0}+\delta\omega$ frequency on the $k^{-}$ laser
beam) while ramping off the RF field. The loading procedure then continues
by slowly ramping on the remaining frequency components on the $k^{-}$
beam, and finally by ramping on the magnetic field gradient (essentially
according in the lattice sites from infinity). This procedure leaves
the BEC in the $q=0$ crystal momentum state in a single Floquet band.

\section{Conclusions}

Initial proposals \cite{Juzeliunas2006,Spielman2009,Gunter2009} and
experiments \cite{Lin2009b} with geometric gauge potentials were
limited by the small spatial regions over which these existed. Here
we described a proposal that overcomes these limitations using laser
coupling reminiscent of a frequency comb: temporally pulsed Raman
coupling. Typically, techniques relying on temporal modulation of
Hamiltonian parameters to engineer lattice parameters suffer from
micro-motion driven heating. Because our method is applied to atoms
initially in free space, with no optical lattice present, there are
no a priori resonant conditions that would otherwise constrains the
modulation frequency to avoid transitions between original Bloch bands
\cite{Weinberg15PRA}.

Still, no technique is without its limitations, and this proposal
does not resolve the second standing problem of Raman coupling techniques:
spontaneous emission process from the Raman lasers. Our new scheme
extends the spatial zone where gauge fields are present by adding
side-bands to Raman lasers, ultimately leading to a $\propto\sqrt{N}$
increase in the required laser power (where $N$ is the number of
frequency tones), and therefore the spontaneous emission rate. As
a practical consequence it is likely that this technique would not
be able reach the low entropies required for many-body topological
matter in alkali systems \cite{Goldman2014}, but straightforward
implementations with single-lasers on alkaline-earth clock transitions
\cite{Fallani16PRL,Kolkowitz2016socSr} are expected to be practical.

\section*{Appendix: Stroboscopic evolution operator}

The stroboscopic evolution operator (\ref{eq:time-evolution}) reads
explicitly

\begin{equation}
U(2\pi,0)=U_{0}U_{\rm{kick}}^{(1)}U_{0}U_{\rm{kick}}^{(0)}=e^{-{\rm i}\pi\left[E_{\rm{r}}\bm{k}^{2}+\frac{1}{2}\sigma_{3}\beta x\right]}e^{-{\rm i}v_{1}(y)}e^{-{\rm i}\pi\left[E_{\rm{r}}\bm{k}^{2}+\frac{1}{2}\sigma_{3}\beta x\right]}e^{-{\rm i}v_{0}(y)}.\label{eq:time-evolution-1}
\end{equation}
In the main we have approximated the evolution operator by Eq. (\ref{eq:effective-time-evolution-op}).
To estimate the validity of the latter equation, let us make use of
the Baker-Campbell-Hausdorff (BCH) formula

\begin{equation}
e^{X}e^{Y}=e^{Z}\quad\mathrm{with}\quad Z=X+Y+\frac{1}{2}\left[X,Y\right]+\ldots\label{eq:BCH}
\end{equation}
and consider this expansion up to the leading term $\frac{1}{2}\left[X,Y\right]$,
essentially the second term in the Magnus expansion. 

Neglecting the commutation between $E_{\rm{r}}\bm{k}^{2}$ and $\sigma_{3}\beta x$
, one can write

\begin{equation}
U_{0}=e^{-{\rm i}\pi\left[E_{\rm{r}}\bm{k}^{2}+\frac{1}{2}\sigma_{3}\beta x\right]}\approx e^{-{\rm i}\pi E_{\rm{r}}\bm{k}^{2}}e^{-i\frac{\pi}{2}\sigma_{3}\beta x}\label{eq:time-evolution-0-1}
\end{equation}
The error in doing so is approximately $-i\frac{\pi}{4}E_{\rm{r}}\beta\sigma_{3}\left[x,\bm{k}^{2}\right]=\frac{\pi}{2}E_{\rm{r}}\beta k_{x}\sigma_{3}$.
Since $E_{\rm{r}}\beta\ll1$, this provides a small momentum shift
along the $x$ direction. Furthermore, we shall neglect the commutation
between $E_{\rm{r}}\bm{k}^{2}$ and $v_{l}(y)$. The error in doing
so is approximately $-i\frac{\pi}{2}E_{\rm{r}}\sigma_{x,y}\left[y,\bm{k}^{2}\right]=\pi E_{\rm{r}}k_{y}\sigma_{3}$.
Since the Floquet frequency $\omega$ greatly exceeds the recoil frequency
$E_{\rm{r}}\ll1$ and $\beta<1$, this also provides a small momentum
shift along the $y$ direction.With these assumptions, one has 
\[
U(2\pi,0)=U_{0}U_{\rm{kick}}^{(1)}U_{0}U_{\rm{kick}}^{(0)}\approx e^{-{\rm i2}\pi E_{\rm{r}}\bm{k}^{2}}e^{-{\rm i}2\pi v_{\rm{eff}}}.
\]
where
\[
e^{-{\rm i}2\pi v_{\rm{eff}}}=e^{-{\rm i}\pi\sigma_{3}\beta x/2}e^{-{\rm i}v_{1}(y)}e^{-{\rm i}\pi\sigma_{3}\beta x/2}e^{-{\rm i}v_{0}(y)}\,.
\]
Finally under the above assumptions one can merge the exponents in
$U(2\pi,0)$, giving Eq.(\ref{eq:effective-time-evolution-op}).

\section*{Acknowledgements}

We thank Immanuel Bloch, Egidijus Anisimovas and Julius Ruseckas for
helpful discussions. This research was supported by the Lithuanian
Research Council (Grant No.\ MIP-086/2015). I.~B.~S.\ was partially
supported by the ARO's Atomtronics MURI, by AFOSR's Quantum Matter
MURI, NIST, and the NSF through the PCF at the JQI.


\end{document}